\documentclass[twocolumn]{aastex7}

\shorttitle{Verification of fast rotating asteroid candidates}
\definecolor{lightcyan}{HTML}{A1FFFF}

\begin{document}

\title{Dense‑photometry validation of super‑fast‑rotating asteroid candidates}

\author[orcid=0000-0001-6349-6881,gname=Bojan, sname='Novakovic']{Bojan Novakovi\'c} 
\affiliation{Department of Astronomy, Faculty of Mathematics, University of Belgrade, Studentski trg 16, 11000 Belgrade, Serbia}
%\affiliation{Studentski trg 16, 11000 Belgrade, Serbia}
\email{bojan@matf.bg.ac.rs}

\author[orcid=0000-0002-7332-6269,gname=Pedro,sname=Gutiérrez]{Pedro J. Gutiérrez}
\affiliation{Instituto de Astrofísica de Andalucía, CSIC, Glorieta de la Astronomía s/n, E-18008 Granada, Spain}
%\affiliation{Glorieta de la Astronomía s/n, E-18008 Granada, Spain}
\email{}

\begin{abstract}

Super–fast rotators (SFRs; $P < 2.2\,$hr) are of great importance in asteroid studies; yet, many reported detections suffer from aliasing caused by an insufficient observation cadence. We present dense CCD photometry for 15 SFRs candidates (14 after excluding 11219 Benbohn, whose published period already exceeds the spin barrier) observed from 2023 Aug 11 to 2024 Aug 11 with the 1.5~m Sierra Nevada and 1.4~m AS Vidojevica telescopes. Our dataset comprises approximately 2,400 calibrated data points, with per-measurement formal errors of $0.02$–$ 0.04$ mag and total on-target coverage of $2$–$ 13$ hr per object. We have reliably determined periods for 9 targets. In terms of spin rate, we have confirmed four SFRs with periods of 1.06–1.84 hr and peak-to-peak amplitudes of 0.054–0.685 mag. Three candidates remain ambiguous, while the rest are reclassified, showing the best solutions with periods greater than 2.5 hr. By extrapolating from our confirmation rate (4/14) to the 3.9$\%$ occurrence rate found in the Light Curve Database (LCDB) yields a central SFR fraction of 1.1$\%$, with a one-sigma lower bound of 0.6$\%$ among kilometre‑scale asteroids.

\end{abstract}



\section{Introduction}

\subsection{Rotation periods of asteroids}

The spin period is an essential parameter for understanding the asteroids, including, for instance, their surface properties \citep{2014Natur.508..233D,2020Icar..33813443S,2024PSJ.....5...11N}, thermophysical modeling \citep{delbo-etal_2015,2018Icar..309..297H,2022PSJ.....3...56H}, internal structures \citep{2014Natur.512..174R,2019Icar..317..354H,2025A&A...695A..30F}, and modeling non-gravitational effects such as Yarkovsky and YORP \citep{2015aste.book..509V,2022SerAJ.204...51F}.

Rotation periods are derived from asteroids' light curves constructed from photometric observations. We can distinguish between dense and sparse photometry depending on the frequency of collected observations. A classical approach produces a dense dataset in which a target is observed continuously over longer periods (for instance, a whole night, or over 2-3 nights). This approach provides the best light curves and the most reliable periods. It is, however, limited to a relatively small number of objects due to the large demands in
telescope time to obtain such data. For these reasons, until recently, the number of asteroids with determined rotation periods
was very limited.

In recent years, different types of sky surveys have provided photometric data for a large number of asteroids, which is also used to derive their spin periods. It significantly increases the available rotation periods and allows a better understanding of some asteroid properties. For instance, it becomes evident that the number of slowly rotating asteroids was
underestimated considerably due to observational biases \citep[e.g.][]{2020ApJS..247...26P,2024A&A...687A.277C}.

The sky surveys, however, produce so-called sparse photometry data. Reducing such data requires additional steps (e.g., corrections for changes in observational geometry and accounting for data collected in different filters). Generally, it makes the extraction of rotation periods more challenging and less reliable \citep[e.g.,][]{2011Icar..216..610W,2024AJ....168..181G}. The exact sampling and the level of the problem depend on each survey cadence. For instance, ground-based telescopic surveys that produce sparse data inevitably have signals in the object’s periodogram, typically at or near 12, 24, 48, and 96 hr — related to the Earth's day-night cycle. These cadences cause aliasing, an effect where peaks in a periodogram are not at the actual period of an observed object. 

In some cases, such as Gaia photometric observations of small solar system bodies, additional complications arise from the fact that the time sampling is neither regular nor random. The spacecraft’s ~6 hr spin and dual fields of view impose quasi-regular windows, but the resulting cadence is far from uniform or strictly periodic \citep[see][Appendix B]{2024A&A...687A.277C}.

Although all surveys collect data less frequently than targeted observations, there are also differences among untargeted observations, depending on the specific cadence used by each survey. Therefore, not all surveys are equally suitable for the construction of asteroid light curves and spin-period determination. Based on this, \citet{2011Icar..216..610W} proposed a limit of 16 observations per night on 16 nights within 30 days, the so-called 16/16 sampling rule, as a distinction between the different classes. Surveys collecting more dense data than the 16/16 sampling rule are obviously more suitable for reliably obtaining asteroids' light curves and determining the periods than surveys collecting the data less frequently. To distinguish between these two classes, we will call the data collected from surveys fulfilling the 16/16 rule semi-dense observations, while the data coming from the other surveys will be classified as sparse data. Therefore, the photometric data used to determine asteroid spin periods can be divided into three categories, namely dense, semi-dense, and sparse. Each category reflects different suitability for constructing asteroid light curves and different levels of reliability in the derived spin periods.

\subsection{Review of existing data}

The Asteroid Lightcurve Database \footnote{\url{https://minplanobs.org/mpinfo/php/lcdb.php}} (LCDB) contains rotation periods for about 36,000 asteroids (the release from October 2023). This represents less than $3\%$ of the population of known asteroids. The database also provides a quality code used to describe the reliability of each period solution. The sample includes about 1,700 objects with highly unreliable period solutions (quality flag $ U\leq 1+$), and about 3,300 lower limit period solutions that are not even given a quality flag. 

\begin{figure}[ht!]
\plotone{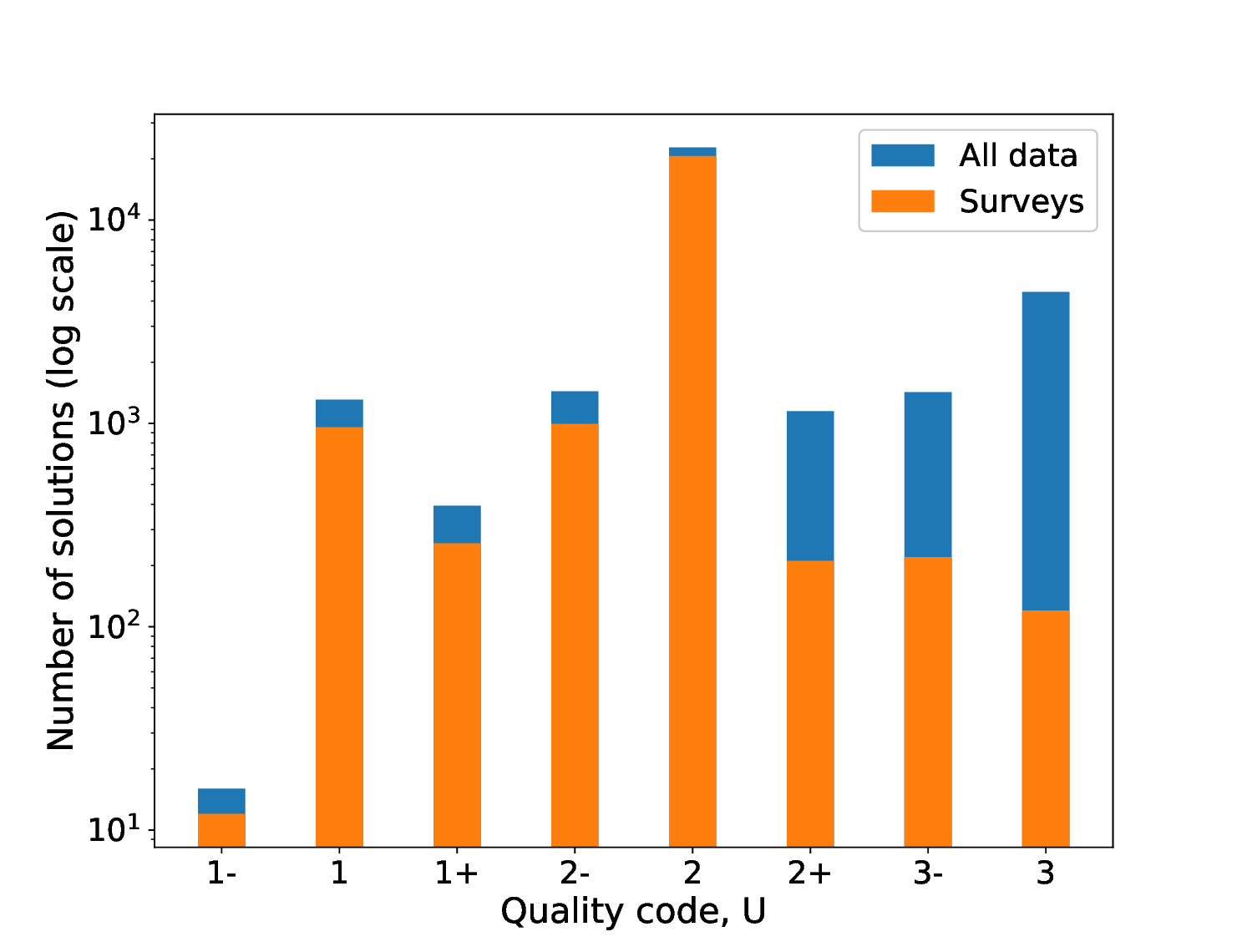}
\caption{The histogram of quality codes for rotational periods of asteroids from the LCDB.}
\label{fig:general}
\end{figure}

Among the remaining about 31,000 asteroids with quality flag $U\geqslant 2-$, more than 70$\%$ have rotation periods determined based on data coming from different sky surveys (see Fig.~\ref{fig:general}). This implies that on one side, the number of asteroids with known spin periods is limited, and on the other side, sky surveys play a more and more important role in their determination.

%% The "ht!" tells LaTeX to put the figure "here" first, at the "top" next
%% and to override the normal way of calculating a float position.
%% The asterisk after "figure" tells the compiler to span multiple columns
%% if a two column style is selected.

With many ongoing surveys and the upcoming Vera C. Rubin Observatory, which will carry out the Legacy Survey of Space and Time (LSST), the potential for the determination of different asteroid parameters, including rotation periods, will increase significantly, and new
data will fill our databases. Therefore, using semi-dense and sparse data will undoubtedly be the standard. 

Still, the role of dense photometry produced in targeted observations will remain important. Despite the survey data outnumbering dense photometry by orders of magnitude, the most reliable spin periods will still be determined based on the targeted observations. Due to this, dense photometry will remain an essential part of the spin state characterisation, either alone or combined with survey data, when high accuracy is required. This is the case, for instance, of near-Earth objects \citep[e.g.][]{2022EM&P..126....6V,2025A&A...695A.139F} or for studying specific groups of asteroids, such as binary asteroids \citep[see, e.g.,][]{2025P&SS..Chiorny}.

Dense photometry also provides essential input for verifying and validating rotation-period solutions obtained from large-scale surveys (sparse data), and it has long served as the benchmark set against which algorithms are tested. In most studies, however, the same dense light curves are also used to tune or calibrate the models, so the apparent agreement partly reflects this built-in bias. An unbiased assessment, therefore, requires new dense light curves for a representative sample of asteroids whose periods have been derived from sparse photometry. Comparing these results with the original sparse-data solutions will reveal the actual level of agreement and, in turn, enhance confidence in rotation periods inferred from survey databases.

\subsection{Importance of super-fast rotators}

\begin{figure}[ht!]
\centerline{\includegraphics[width=1.2\columnwidth,angle=-90]{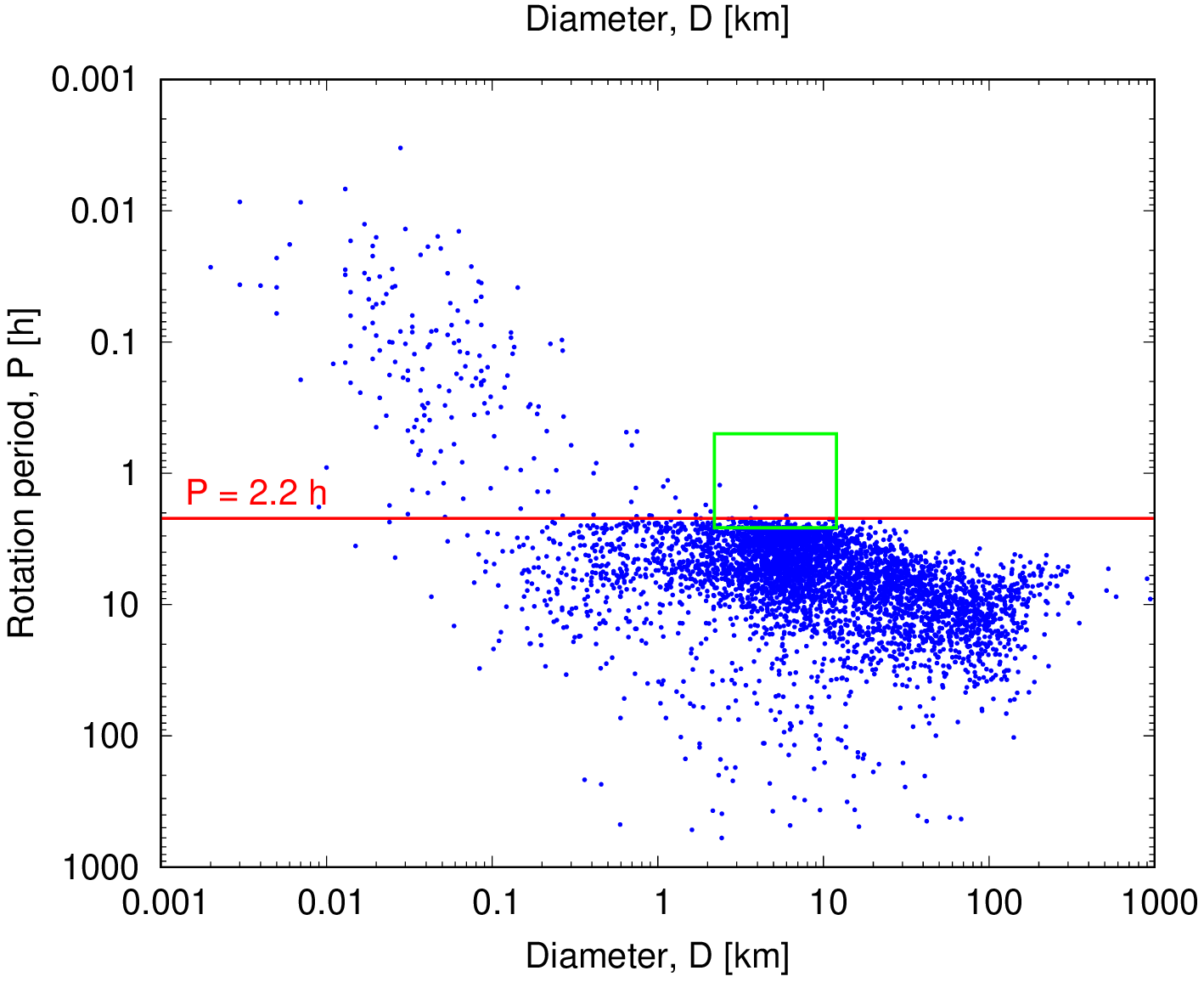}}
\caption{Spin period vs diameter of asteroids from the LCDB. In the upper panel, all data is shown regardless of quality code. The orange points mark periods with $U\geqslant 2-$, while the black dots are objects with the lower quality code. The lower panel shows only data with highly reliable spin periods (blue dots), including less than 4,500 asteroids having a quality code $U=3$. In both panels, the green rectangle shows the intervals where targets of this work are located based on their available spin periods. The horizontal red lines show the spin barrier at P = 2.2 hr.}
\label{fig:PvsD}
\end{figure}

Around the turn of the century, \citet{2000Icar..148...12P} analyzed the distribution of asteroid spin rates vs. size. The authors found that asteroids do not rotate below a spin limit of the cohesionless body, which is approximately 2.2 hr (see also Fig.~\ref{fig:PvsD}). However, with the accumulation of spin period samples, asteroids that rotate with periods less than 2.2 h have been discovered. Furthermore, the limit depends on the asteroid's density, and it is shifted towards larger periods for smaller densities \citep[see e.g.][]{2025A&A...693A..66V,2025ApJ...986L..33T}. These objects are commonly referred to as super-fast rotators (SFRs). As discussed above, asteroid rotation periods are important in many
different aspects. The SFRs put all these to limits.

There are currently hundreds of asteroids known to rotate faster than the spin barrier limit. The list includes the detection of very small \citep[][]{2024MNRAS.527.1633B} and extremely fast rotators \citep[][]{2024A&A...689A.211D}.
Initially, the identified SFRs were small, measuring less than approximately 150 meters in size. That is in line with the hypothesis suggesting that while objects in the size range between about 150 m and 10 km have a rubble-pile-like internal structure \citep{2018ARA&A..56..593W}, smaller asteroids are primarily monolithic objects \citep{2000Icar..148...12P,2013MPBu...40...42P}, with strong internal forces that hold the body intact even at fast spin rates. Even in that case, the fast spin may cause resurfacing and shedding of surface regolith \citep{2024A&A...684A.172D}.

However, to date, also some large objects with a diameter D $>$ 150 m have been confirmed as SFRs, while several tens of other large objects are regarded as SFRs candidates, according to the LCDB database (see Fig.~\ref{fig:PvsD}). Since gravity alone cannot maintain their structure, this raises questions about whether they could also be monolithic objects, or contain internal forces strong enough to prevent them from disrupting even at spin rates above the "spin barrier." Some indirect evidence shows that at least some of them can be rubble piles with certain tensile strength among the components \citep[e.g.][]{2021MNRAS.502.5277H}, but many mysteries remain.

As an example, the low density of a km-sized asteroid (29075) 1950 DA measured by
Yarkovsky orbital drift and thermal-infrared observations show that it is probably a rubble pile,
and the 2.1216 h spin period requires a minimum cohesion of 44–76 Pa \citep{2014Natur.512..174R}.
However, there are still too few similar results to draw any general conclusions, and it is, therefore, essential to identify new SFRs and reliably derive their cohesion level.

At the same time, the purely monolithic nature of objects below 150 meters in size may also be questionable. In this respect, two rapidly rotating near-Earth asteroids, smaller than 100 meters in size, have recently been found to have low
thermal inertia \citep{2021A&A...647A..61F,2023A&A...675A.134F}, which apparently contradicts their expected
monolithic structure. Such fast spin rates, coupled with the very low gravitational environment
due to the small size, are expected to cause the ejection of material from the surface, thus
preventing regolith grains from being retained. So, what is the reason behind the low thermal
inertia in these bodies? Could they be rubble-pile-like objects with a high porosity? Or could microporosity within the material itself be responsible for the reduced thermal conductivity?

Though the precise answers to the above questions are still missing, these findings blur the border between monoliths and rubble-pile objects. This opens many questions about fast rotators, making their study a hot topic in asteroid-related research.

\subsection{Goals of this work}

Nearly all securely identified SFRs have
diameters $D \lesssim 150$ m, and although several kilometer-sized candidates
have been reported, only a handful satisfy stringent reliability criteria
(Fig.~\ref{fig:PvsD}).  

Detecting such rapid rotation in all-sky surveys is intrinsically challenging
because the time between consecutive measurements is usually days to months,
vastly exceeding the sub-hour periods of interest
\citep[e.g.][]{2018AJ....156..241H, 2023A&C....4400711K}.  As a result, SFRs are
normally uncovered either (i) through dedicated dense photometry of individual
targets \citep[e.g.][]{2020MNRAS.495.3990M, 2020PASP..132f5001R,
2023MNRAS.521.3784L} or (ii) in specialized fast-cadence surveys
\citep[e.g.][]{2019ApJS..241....6C, 2020AJ....160...73Y,
2024AJ....168..184S}.  Both approaches sample only a small fraction of the
population, so the census of SFRs—especially the larger ones—remains
incomplete.  Indeed, \citet{2024AJ....168..184S} estimate that SFRs account for
no more than 0.5\% of main-belt asteroids, underscoring the need for rigorous
verification of every candidate period \citep{2017ApJ...840L..22C}.

Our monitoring program tackles this problem through two objectives:
(i) Assess the reliability of periods derived from sparse or semi-dense survey photometry, and
(ii) Securely confirm (or refute) the super-fast rotation of selected candidates.

For objects whose rapid rotation we confirm, the robust period measurements will permit detailed 
investigations of their internal strength and the level of cohesion required to avert rotational disruption.

Here we report the first results for a sample of 15 SFR
candidates drawn from the catalogs of \citet{2014ApJ...788...17C, 2018ApJS..237...19E, 2019ApJS..242...15E,
2020ApJS..247...26P, 2022ApJ...932L...5C}, none of which previously possessed independent dense light curves.

\section{Methods}

\subsection{Target selection and photometric characteristics of source surveys}
\label{sec:survey_descr}

\begin{deluxetable*}{lcccccccc}
\tablecaption{The data on target asteroids\label{tab:asteroid_data}}
\tablehead{
\colhead{Asteroid} & 
\colhead{$H$\tablenotemark{a}} & 
\colhead{$D$ [km]\tablenotemark{b}} & 
\colhead{$p_V$\tablenotemark{c}} & 
\colhead{$Fam.$\tablenotemark{d}} &
\colhead{$P$ [hr]\tablenotemark{e}} & 
\colhead{$U$\tablenotemark{f}} & 
\colhead{Survey} &
\colhead{$Ref$\tablenotemark{g}} 
}
\startdata
2024 McLaughlin    & 13.04 & 7.915 & 0.173 & Vesta & 1.15    & -- & KMTNet &  \citet{2018ApJS..237...19E,2019ApJS..242...15E} \\
6689 Floss         & 15.05 & 2.994 & 0.164 & -- & 0.88    & -- & KMTNet & \citet{2018ApJS..237...19E,2019ApJS..242...15E} \\ 
9150 Zavolokin     & 13.66 & 5.223 & 0.310 & -- & 1.85391 & 2  & TESS &  \citet{2020ApJS..247...26P} \\
9664 Brueghel      & 13.43 & 11.686 & 0.068 & Themis & 1.15   & -- & KMTNet &  \citet{2018ApJS..237...19E,2019ApJS..242...15E} \\
10461 Dawilliams   & 14.33 & 5.542 & 0.144 & -- & 0.97    & -- & KMTNet &  \citet{2019ApJS..242...15E} \\
11219 Benbohn      & 14.82 & 3.436 & 0.180 & -- & 2.57    & 1  & PTF &  \citet{2014ApJ...788...17C} \\
12549 (1998 QO26)  & 13.40 & 10.777 & 0.088 & -- & 1.13   & -- & KMTNet &  \citet{2019ApJS..242...15E} \\
12768 (1994 EQ1)   & 14.47 & 3.909 & 0.290 & -- & 1.1     & -- & KMTNet &  \citet{2018ApJS..237...19E,2019ApJS..242...15E} \\ 
20546 (1999 RA105) & 13.30 & 11.436 & 0.085 & Hygiea & 0.48   & -- & KMTNet &  \citet{2018ApJS..237...19E,2019ApJS..242...15E} \\
26374 (1999 CP106) & 14.23 & --    & --    & -- & 1.89023 & 2  & TESS &  \citet{2020ApJS..247...26P} \\ 
29210 Robertbrown  & 13.88 & 9.077 & 0.094 & -- & 2       & -- & KMTNet &  \citet{2018ApJS..237...19E,2019ApJS..242...15E} \\ 
29930 (1999 JT41)  & 13.99 & --    & --    & -- & 1.29628 & 2  & TESS &  \citet{2020ApJS..247...26P} \\
31029 (1996 HC16)  & 15.28 & 2.388 & 0.491 & -- & 1.8     & 2+ & ZTF &  \citet{2022ApJ...932L...5C} \\
44145 (1998 HJ101) & 15.86 & --    & --    & -- & 1.74    & 3  & ZTF &  \citet{2022ApJ...932L...5C} \\
85128 (1979 HA)    & 13.84 & 5.299 & 0.275 & Phocaea & 1.65369 & 2  & TESS &  \citet{2020ApJS..247...26P} \\
\enddata
\tablenotetext{a}{Absolute magnitude (from JPL).}
\tablenotetext{b}{Diameter (in km), from the JPL.}
\tablenotetext{c}{Geometric albedo.}
\tablenotetext{c}{Family membership (from Asteroid Families Portal; \citet{2022CeMDA.134...34N})}
\tablenotetext{e}{Rotation period previously reported in literature.}
\tablenotetext{f}{Quality code from the Asteroid Lightcurve Database (LCDB).}
\tablenotetext{g}{Source of the period solution.}
\end{deluxetable*}

The available rotation periods of our targets vary in reliability and are based on four data sets, including space–based and targeted wide–field, fast–cadence ground observations. Eight of the targets are selected from \citet{2018ApJS..237...19E,2019ApJS..242...15E}, who used the Korea Microlensing Telescope Network (KMTNet), four are from \citet{2020ApJS..247...26P}, and are based on the Transiting Exoplanet Survey Satellite (TESS) data, two are from \citet{2022ApJ...932L...5C} who used Zwicky Transient Facility (ZTF), and one from \citet{2014ApJ...788...17C}, which are derived from the Palomar Transient Factory (PTF). Below, we summarize the most relevant properties for rotational–period studies of each survey.

%-------------------------------------------------------------------

\paragraph{Korea Microlensing Telescope Network (KMTNet).} 
KMTNet employs three identical 1.6\,m telescopes; our data come from the South African node. Each telescope features a $2^\circ\times2^\circ$ field of view, using four $9\mathrm{k}\times9\mathrm{k}$ CCDs at 0.40\,arcsec/pixel. Observations utilized 60-second exposures, achieving an effective cadence of about 135 seconds. Between October 2016 and February 2017, newly discovered near-Earth asteroids were observed within 4--44~days of discovery, cycling Johnson–Cousins $V$, $R$, and $I$ filters (sequence VRVI). This targeted rapid-cadence approach is optimal for confirming asteroid rotation periods, particularly among fast-rotating candidates.

\paragraph{Transiting Exoplanet Survey Satellite (TESS).} 
TESS acquires continuous full-frame images every 30 minutes, covering sectors for approximately 28 days each. This uninterrupted cadence avoids the diurnal aliasing common in ground-based observations, providing highly reliable asteroid rotation periods. Although periods shorter than about 1 hour cannot be resolved due to Nyquist limits, TESS effectively detects super-fast rotators with periods in the range 1--2.2 hr.

\paragraph{Zwicky Transient Facility (ZTF).} ZTF, the successor of PTF, uses the same 48-inch telescope equipped with a 576-megapixel camera (47 deg$^2$ field of view). It observes in $g$, $r$, and $i$ filters, typically achieving a 5$\sigma$ limiting magnitude of $\sim$20.4 mag ($r$ band) with 30-second exposures. The asteroid data used by \citet{2022ApJ...932L...5C} are from two high-cadence $r$-band surveys in early and late 2019, covering 1692 deg$^2$ around the ecliptic near opposition. Observations were conducted at a cadence of approximately 5 minutes and with nightly durations of up to 7 hr.

\paragraph{Palomar Transient Factory (PTF).} PTF utilized the 48-inch Samuel Oschin Telescope with an 11-chip mosaic CCD (field of view 7.26 deg$^2$, pixel scale 1.0\arcsec/pixel), observing primarily in the $R$ band. The asteroid rotation-period data stem from the 10k Asteroid Rotation Periods (10kARPs) campaign, conducted on 15--18 February 2013, covering 87 deg$^2$ along the ecliptic plane with a 20-minute cadence and 60-second exposures.

Together, these surveys provide broad coverage in magnitude, cadence, and ecliptic longitude, enabling us to assemble a diverse sample of 15 super-fast-rotator candidates for follow-up dense photometry. We note that periods from \citet{2018ApJS..237...19E,2019ApJS..242...15E} are only lower limits, and therefore these are not graded according to the LCDB 's quality flags (see Table~\ref{tab:data}).

%%\pagecolor{blue!5}

\subsection{The equipment used for observations}

The observations presented in this work have been collected from the Sierra Nevada Observatory and Astronomical Station of Vidojevica. The information about the sites and equipment is provided in Table~\ref{tab:equ}.

\begin{deluxetable}{lll}
\tablecaption{The equipment used for observations.\label{tab:equ}}
\tablehead{
\colhead{Parameter} & \colhead{Sierra Nevada Observatory} & \colhead{AS Vidojevica}
}
\startdata
\multicolumn{3}{c}{\textbf{Site Information}} \\
Acronym        & SNO                & ASV \\
MPC code       & J86                & C89 \\
Latitude       & N 37$^\circ$ 03$'$ 51$''$ & N 43$^\circ$ 08$'$ 24.6$''$ \\
Longitude      & W 03$^\circ$ 23$'$ 05$''$ & E 21$^\circ$ 33$'$ 20.4$''$ \\
Altitude       & 2896 m             & 1150 m \\
\multicolumn{3}{c}{\textbf{Telescope and Detector}} \\
Telescope      & 1.5 m, f/8          & 1.4 m, f/8 \\
CCD Model      & Andor iKon-L 936     & Andor iKon-L 936 \\
Array Size     & 2048 $\times$ 2048 pixels & 2048 $\times$ 2048 pixels \\
Pixel Size     & 13.5 $\mu$m $\times$ 13.5 $\mu$m & 13.5 $\mu$m $\times$ 13.5 $\mu$m \\
Field of View (FOV) & $7.8' \times 7.8'$ & $13.3' \times 13.3'$ \\
Pixel Scale    & $0.23''$/pixel & $0.39''$/pixel \\
\multicolumn{3}{c}{\textbf{Observing Setup}} \\
Filter Band      & $R$                 & $R$ \\
Binning          &  2$\times$2 & 2$\times$2 \\
\enddata
\end{deluxetable}

\subsection{Data reduction and period search}

The image processing and measurement were done using procedures incorporated into the Tycho Tracker software \citep{2020JAVSO..48..262P}. All raw images underwent bias and flat-field corrections (ASV images were also dark-corrected). Aperture photometry was calibrated against the ATLAS All-Sky Stellar Catalog \citep{2018ApJ...867..105T}. Observing circumstances are listed in Table~\ref{tab:obs}.

For period analysis, we used our Python implementation of the FALC algorithm \citep{1989Icar...77..171H} and fitted the 2nd–6th order Fourier series. The search was performed using 10,000 steps and initially spanned a range from 0.1 to 50 hr. However, this was then repeated in a shorter range, depending on the initial results.

Uncertainties are derived from resampling by applying the fitting for various periods within a given range. For each trial period, we compute the goodness of fit, $\chi^{2}(P)$, on a sampled grid that spans the search range. The nominal period, $P$, corresponds to the global minimum $\chi^{2}_{\min}$. To quantify its formal $1\sigma$ uncertainty we adopt the standard \mbox{$\Delta\chi^{2}=1$} criterion for a single parameter of interest: all periods that satisfy $\chi^{2}(P)\le\chi^{2}_{\min}+1$ belong to the $68\%$ confidence interval. The half-width of this interval is quoted as the period error, $\sigma_{P}$.

Once the Fourier model is fixed, the peak-to-peak amplitude follows
directly from its maximum and minimum. To evaluate how sensitive that amplitude is to uncertainties in the fitted coefficients, we generate an ensemble of $N=500$ surrogate curves. Each curve derives its coefficients from the multivariate normal distribution defined by the best-fit parameters and their covariance matrix, as returned by the least-squares routine. The amplitudes of these surrogate curves form a narrow distribution whose standard deviation is taken as the formal $1\sigma$ uncertainty, $\sigma_{\mathrm{amp}}$. This Monte-Carlo procedure naturally incorporates correlations between coefficients and the non-linear way in which they combine to yield the final peak-to-peak value.

%%\movetabledown=40mm

%%\begin{rotatetable*}
\begin{deluxetable*}{lcccccccr}
\tablecaption{Observational circumstances \label{tab:obs}}
\tablehead{
\colhead{Asteroid} & \colhead{UT Date} & \colhead{Observatory} & \colhead{Exp. [s]} & \colhead{\# of images} & \colhead{Seeing ['']} & \colhead{$r_h$ [au]} & \colhead{$\Delta$ [au]} & \colhead{$\alpha$ [deg]} 
}
\startdata
2024 McLaughlin   & 23-01-2024  & OSN &  90 & 106 &  2.1 & 2.173  & 1.769 & 26.5   \\
6689 Floss        & 03-05-2024  & OSN &  50 & 207 &  2.1 & 1.950  & 0.944 &  2.8   \\
9150 Zavolokin    & 06-01-2024  & OSN & 150 & 125 &  2.6 & 2.723  & 1.808 &  9.0   \\
9664 Brueghel     & 04-03-2024  & OSN & 120 &  64 &  2.9 & 2.891  & 1.906 &  2.9  \\
10461 Dawilliams  & 11-08-2023  & ASV & 120 &  78 &  1.8 & 2.086  & 1.266 & 21.1  \\
10461 Dawilliams  & 12-08-2023  & ASV & 120 & 151 &  1.5 & 2.086  & 1.257 & 20.8  \\
10461 Dawilliams  & 13-08-2023  & ASV & 120 & 145 &  1.5 & 2.085  & 1.249 & 20.4 \\
11219 Benbohn     & 03-02-2024  & OSN &  90 & 195 &  2.5 & 2.212  & 1.227 &  1.0 \\
11219 Benbohn     & 04-02-2024  & OSN &  90 & 103 &  2.4 & 2.210  & 1.224 &  0.5  \\
12549 (1998 QO26) & 02-02-2024  & OSN & 120 &  88 &  2.2 & 2.836  & 1.890 &  6.9  \\
12549 (1998 QO26) & 04-02-2024  & OSN & 120 & 139 &  2.8 & 2.837  & 1.883 &  6.2  \\
12768 (1994 EQ1)  & 07-01-2024  & OSN & 180 &  97 &  2.9 & 2.062  & 1.387 & 24.4  \\
20546 (1999 RA105) & 03-03-2024 & OSN &  90 & 106 &  2.5 & 2.806  & 1.826 &  3.8  \\
26374 (1999 CP106) & 14-04-2024 & OSN & 180 &  42 &  3.3 & 2.657  & 1.660 &  3.2  \\
29210 Robertbrown & 10-08-2024  & ASV & 150 &  75 &  1.9 & 2.667  & 1.655 &  1.8  \\
29210 Robertbrown & 11-08-2024  & ASV & 160 &  88 &  1.8 & 2.665  & 1.653 &  1.3  \\
29930 (1999 JT41) & 16-02-2024  & OSN &  90 & 104 &  3.4 & 2.546  & 1.594 &  7.5  \\
31029 (1996 HC16) & 20-01-2024  & OSN &  90 & 218 &  2.0 & 2.083  & 1.115 &  6.5  \\
44145 (1998 HJ101) & 10-08-2024 & ASV & 150 &  47 &  1.8 & 1.803  & 0.813 & 10.3  \\
44145 (1998 HJ101) & 11-08-2024 & ASV & 150 &  43 &  1.6 & 1.803  & 0.811 &  9.7  \\
85128 (1979 HA)   & 21-01-2024  & OSN & 120 & 222 &  3.5 & 2.200  & 1.414 & 19.3  \\
\enddata
\end{deluxetable*}
%%\end{rotatetable*}

\section{Results}

\begin{deluxetable}{lrcc}
\tablecaption{The derived spin parameters of target asteroids \label{tab:data}}
\tablehead{
\colhead{Asteroid}  & \colhead{Period [hr]} & \colhead{Amplitude} & \colhead{St.\tablenotemark{a}} 
}
\startdata
2024 McLaughlin    &  1.213 $\pm$ 0.003 & 0.054 $\pm$ 0.006 & C \\
6689 Floss         &  3.848 $\pm$ 0.666 & 0.368 $\pm$ 0.011 & R \\ 
9150 Zavolokin     &  1.842 $\pm$ 0.010 & 0.194 $\pm$ 0.009 & A \\
9664 Brueghel      &  1.058 $\pm$ 0.976 & 0.043 $\pm$ 0.010 & A \\
10461 Dawilliams   &  2.725 $\pm$ 0.003 & 0.386 $\pm$ 0.003 & R \\
11219 Benbohn      &  2.555 $\pm$ 0.003 & 0.175 $\pm$ 0.004 & - \\
12549 (1998 QO26)  &  70.8  $\pm$ 1.6   & 1.9   $\pm$ 0.3   & R \\
12768 (1994 EQ1)   &  6.228 $\pm$ 0.018 & 0.467 $\pm$ 0.007 & R \\ 
20546 (1999 RA105) &  5.1   $\pm$ 0.1   & 0.344 $\pm$ 0.007 & R \\ 
26374 (1999 CP106) &  4.84  $\pm$ 0.03  & 0.336 $\pm$ 0.023 & A \\ 
29210 Robertbrown  &  3.469 $\pm$ 0.003 & 0.126 $\pm$ 0.011 & R \\ 
29930 (1999 JT41)  &  1.393 $\pm$ 0.010 & 0.128 $\pm$ 0.007 & C \\
31029 (1996 HC16)  &  1.797 $\pm$ 0.003 & 0.190 $\pm$ 0.005 & C \\
44145 (1998 HJ101) &  1.737 $\pm$ 0.003 & 0.685 $\pm$ 0.008 & C \\
85128 (1979 HA)    &  3.319 $\pm$ 0.010 & 0.136 $\pm$ 0.003 & R \\
\enddata
\tablenotetext{a}{Status as the SFRs: C - confirmed, A - ambiguous, R - rejected}
\end{deluxetable}

\subsection{Individual objects}

\textbf{2024 McLaughlin} was observed on the night of 23rd January 2024, for about 2.7 hr. The raw plot shows maximum variations of 0.2 mag with no obvious rotational-related variations. Although formal errors of individual data points are typically low, below 0.02 mag, real uncertainties may be higher, making it difficult to reliably fit the period due to the relatively low amplitude.
Nevertheless, we fit the solution using the 4th-order FALC and find the best-fit solution of 1.213 hr with an amplitude of 0.054 mag. This is in good agreement with the result obtained by \citet{2018ApJS..237...19E,2019ApJS..242...15E}, who suggested a lower limit period solution of 1.15 hr.
Still, the scattering of the data points is significant compared to the amplitude of variations, and therefore, further observations are needed to derive a reliable rotation period solution.

\textbf{6689 Floss} was also observed on one night, for a total of about 3 hr. The resulting light curve is well-defined, but the period is somewhat longer than the interval covered by the observations. The best-fit solution suggests 3.85 hr period and an amplitude of about 0.37 mag. It is significantly longer than the minimum period of ~0.9 hr proposed by \citet{2018ApJS..237...19E,2019ApJS..242...15E}. Although our solution is uncertain for some 20$\%$, it rules out periods shorter than 2 hr; therefore, asteroid (6689) Floss is not a super-fast-rotating object.

\textbf{9150 Zavolokin} was observed for a single night (~5 hr) from SNO. Two TESS-based solutions in the literature suggest
periods of 1.85 hr \citep{2020ApJS..247...26P} and 3.71 hr \citep{2025A&A...693A..66V}. The FALC fit to our data provides a reasonable approximation of the asteroid's light curve, yielding a rotational period of approximately 1.842 hr. However, outliers and a pronounced light-curve asymmetry (including a less prominent second maximum) limit the fit quality and introduce ambiguity, reducing confidence in the derived spin parameters. Although it could be a super-fast rotator, the observed discrepancies indicate that multi-night photometry is needed for a more definitive characterization of its rotational properties.

\textbf{9664 Brueghel} was observed for slightly over 2 hr. Similarly to (2024) McLaughlin, the light curve is not well defined due to the relatively small amplitude of variations, and somewhat larger scattering of data points, possibly due to bad seeing. Therefore, we can not derive a highly reliable period solution in this case. Nevertheless, the best-fit solution indicates a period of about 1.06$\pm$0.98 hr, which is slightly shorter but still in reasonable agreement with the lower-limit period of 1.15 found by \citet{2018ApJS..237...19E,2019ApJS..242...15E}. Further observations are needed to verify the period, but (9664) Brueghel could be a super-fast-rotating asteroid. The obtained small light curve amplitude of only 0.043 mag further strengthens this possibility.

\textbf{10461 Dawilliams} was the best observed object in this campaign. It is observed on three nights from ASV. On August 11th, 2023, observations were conducted for almost 2.5 hr, divided into two runs separated by 2 hr due to adverse weather conditions. Observations were also collected on August 12th and 13th, 2023, for 5.2 and 5 hr, respectively, without gaps. Joint observations from those three nights yielded a well-defined light curve, enabling the reliable determination of a rotation period. As the best-fit solution, we found a period of 2.725 hr and a corresponding amplitude of 0.386 mag, obtained by a 5th-degree fit. The obtained period is about 3 times longer than the lower limit found by \citet{2018ApJS..237...19E,2019ApJS..242...15E}. It also indicates that this asteroid is not a super-fast rotator, though it spins only about 0.5 hr slower than the spin barrier.

\textbf{11219 Benbohn} was observed on two nights, on 2nd and 4th February 2024, for about 5 and 2.5 hr, respectively. Despite some noise in the data, the resulting period solution appears to be reliably determined. The best-fit solution gives a period of 2.555 hr. Our result is entirely in agreement with the solution of 2.57 hr, previously found by \citep{2014ApJ...788...17C}.

\textbf{12549 (1998 QO26)} was also observed on two nights, on 2nd and 4th February 2024, for about 2.9 and 4.7 hr, respectively. There is also a 1-hour gap in the data from the first night, due to an asteroid passing near a star that was already at saturation. Nevertheless, if the real period is close to the lower limit of 1.13 hr \citep{2018ApJS..237...19E,2019ApJS..242...15E}, it should be visible in our data. Still, our data reveals a constantly increasing brightness on the first night and a continually decreasing brightness on the second night. Therefore, the actual spin period could be significantly longer than approximately 1 hour. Recently, \citet{2024A&A...687A.277C} also estimated the period of this asteroid from Gaia data \citep{2023A&A...674A..12T}. The authors found a period of about 67.2 hr. If so, our data is not enough to reliably estimate the rotation period. All our attempts yield periods above 8 hr. Limiting our search between 50 and 80 hr, we obtained a solution of about 70.8 hr and an amplitude of 1.9 mag. In any case, the (12549) 1998~QO26 is not a super-fast rotator.

\textbf{12768 (1994 EQ1)} was observed on 7th January 2024 night for about 5 hr. The light curve is well-defined, but it appears to cover an interval slightly shorter than the rotational period. The best-fit solutions indicate a period of about 6.228 hr and an amplitude of 0.467 mag. Though additional data is needed to get a very accurate period, the obtained results rule out super-fast rotation.

\textbf{20546 (1999 RA105)} was observed on 5 March 2024 night. A light curve is well-defined; however, the approximately 2.7-hour observing window did not encompass the entire rotation period. A single-peaked light curve yields the best fit, indicating a period of approximately 2.54 hr. However, as such a light curve is likely representing only half a period, for a nominal solution, we adopted 5.1 hr. Both periods are significantly longer than the minimum period of 0.5 hr suggested by \citet{2018ApJS..237...19E}, and rule out super-fast rotation also for this object.

\textbf{26374 (1999 CP106)} was observed for about 2.1 hr on 14th April 2024 night. The resulting light curve is relatively well-defined, but it seems to cover only half of a real rotational period. As the best-fit orbit solution, we found 2.42 hr. This is somewhat longer than the 1.89 hr suggested by \citet{2020ApJS..247...26P}. However, we note that the obtained solution is single-peaked, and the obtained period could be half of the actual period, which would be approximately 4.84 hr. \citet{2025A&A...693A..66V} found a period of 3.78 hr. Our results do not support this solution, but given that our observations covered only slightly more than 2 hr, we cannot reject it. Therefore, further study of this object is needed.

\textbf{29210 Robertbrown} was observed on two nights. On August 10th and 11th, 2024, for a total of 3.2 and 3.9 hr, respectively. The collected data allow for a reliable period determination of about 3.5 hr. Therefore, it is another candidate ruled out as a super-fast rotator.

\textbf{29930 (1999 JT41)} was observed for about 2.7 hr on February 16th, 2024 night. The results are, however, obtained under difficult weather conditions, particularly with poor seeing. Also, the last 25 images are taken slightly below 30 degrees above the local horizon. Therefore, the resulting data points are scattered, and the light curve is not well-defined. Still, thanks to an amplitude of about 0.2 mag, brightness variations related to rotation are visible. The best-fit double-peaked period solution that we found is about 1.393 hr and an amplitude of 0.128 mag. This is slightly longer, but consistent with a period of about 1.3 hr found by \citet{2020ApJS..247...26P}. Therefore, we conclude that this asteroid could be a super-fast rotator. Nevertheless, some further verification is still welcome.

\textbf{31029 (1996 HC16)} was observed for about 5.5 hr on 20th January 2024 night. Despite some scattering in our measurements, the light curve is well-defined. Our best-fit period solution confirms a period of about 1.8 hr found by \citet{2022ApJ...932L...5C}. Therefore, this asteroid should be considered a super-fast rotator.

\begin{figure*}[ht!]
\centerline{\includegraphics[width=2.0\columnwidth,angle=0]{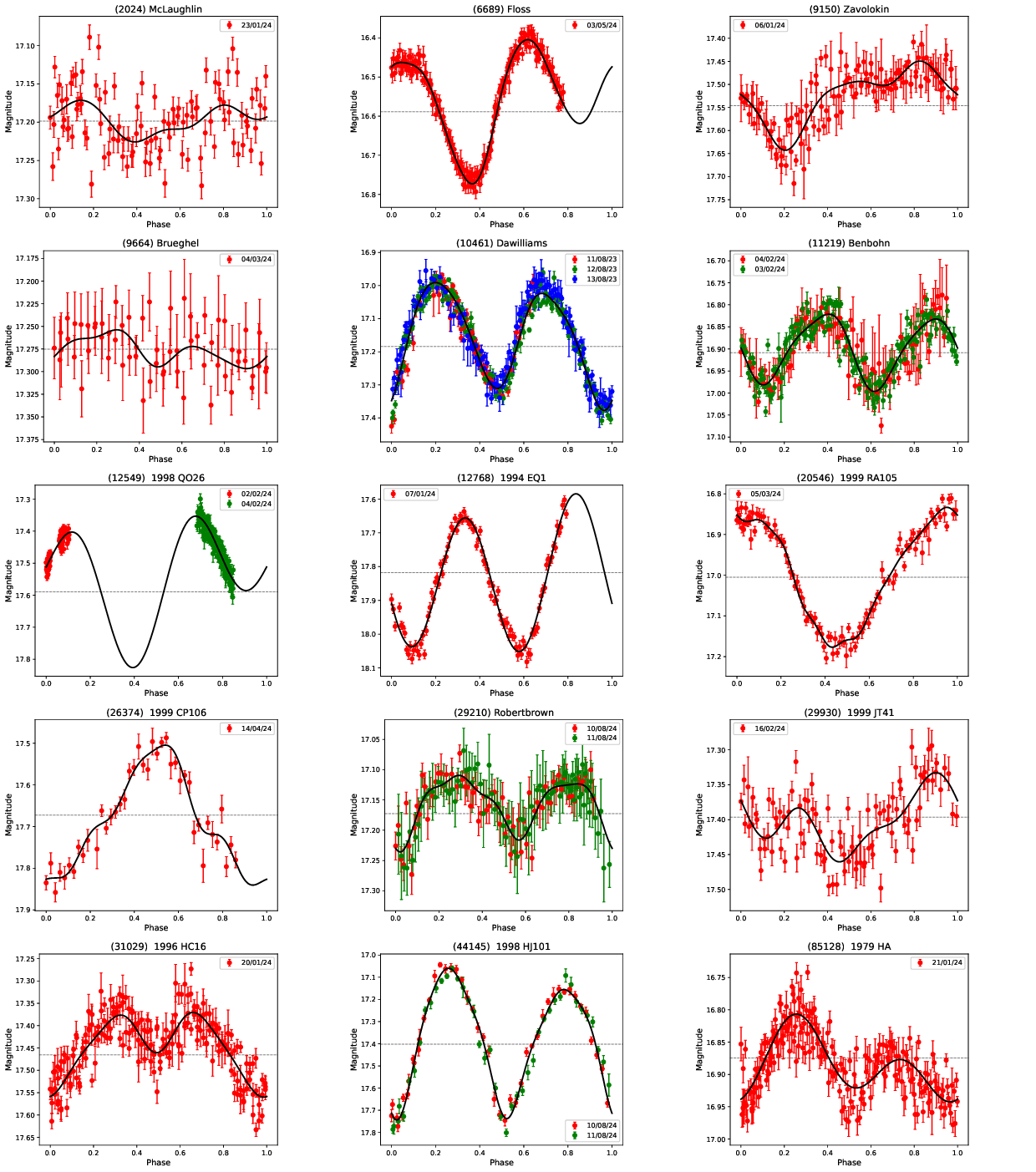}}
\caption{The light curves of 15 asteroids observed in our campaign and the corresponding FALC fits of the data. The horizontal dashed lines mark the mid-points of the light curves.}
\label{fig:lightcurves}
\end{figure*}

\textbf{44145 (1998 HJ101)} was observed on two nights. On August 10th and 11th, 2024, for a total of 2 and 1.8 hr, respectively. \citet{2022ApJ...932L...5C} found a period of 1.74 hr, using ZTF survey data. Our analysis confirms this result, as we found a period of 1.737 hr. Therefore, the asteroid (44145) 1998~HJ101 is a super-fast rotator. Interestingly, the corresponding light curve has a relatively large amplitude of 0.685 mag (see Fig.~\ref{fig:lightcurves}), suggesting an elongated shape.

\textbf{85128 (1979 HA)} was observed for about 7.9 hr on January 21st 2024 night, under poor seeing, especially in the first hour of the observations. Still, the light curve is reasonably well-defined, with two clear peaks but with quite different amplitudes. Our nominal rotation period solution is about 3.32 hr. It is a factor of 2 longer than the solution of 1.654 hr proposed by \citet{2020ApJS..247...26P}, who, we believe, fitted half of the real period. This implies that this object is also not a super-fast rotator.

\subsection{Population fraction of super–fast rotators}\label{sec:population}

Of the 27,157 kilometre‑scale asteroids in the LCDB, 1,048 (3.9$\%$) list nominal periods shorter than 2.2~hr spin barrier. Excluding (11219) Benbohn, whose published period already exceeds the barrier, our dense follow‑up campaign observed fourteen viable fast‑rotator candidates. We confirm four, reclassify seven as slow rotators, and leave three ambiguous. So, the conservative confirmation efficiency is 
$\epsilon = 4/14 = 0.29$. With $n=14$ trials, the binomial $1\sigma$ uncertainty, based on the Wald normal‑approximation \citep[e.g.][]{2011PASA...28..128C}, is $\sigma_\epsilon = \sqrt{\epsilon(1-\epsilon)/n}=0.12$. Multiplying by the LCDB candidate fraction gives the debiased SFR fraction $f_{\mathrm{SFR}} = f_{\mathrm{cand}} \times \epsilon = 0.039 \times 0.29 = 0.011 \; (\text{or } 1.1\%$), with $1\sigma$ error $\sigma_f = f_{\mathrm{cand}}\times\sigma_\epsilon \simeq 0.5\%$. Hence the $1\sigma$ interval is $0.6\% \le f_{\mathrm{SFR}} \le 1.6\%$.

If all three ambiguous targets are eventually confirmed as SFRs, the efficiency would become $\epsilon_{\max}=7/14=0.50$; the corresponding fraction is $f_{\mathrm{SFR,max}} = 0.039\times0.50 = 1.9\%$ with the same binomial error of $\pm 0.5\%$.

Alternatively, restricting the calculation to the eleven targets with secure light‑curve periods yields $\epsilon' = 4/11 = 0.36$, so $ f'_{\mathrm{SFR}} = 0.039\times0.36\approx 1.4\%$ with a comparable $1\sigma$ error of $\pm0.6\%$.

Both estimates are consistent with the $0.4-0.8\%$ SFR fractions reported by recent fast-cadence surveys \citep[e.g.,][]{2024AJ....168..184S,2019ApJS..241....6C}, indicating statistical consistency once the uncertainties are taken into account.

\section{Conclusions}

We obtained dense light curves for 15 asteroids whose literature periods had suggested super–fast rotation.
For 9 objects, the data allowed a reliable determination of the rotation period. Our analysis yields the following main points:

\begin{enumerate}
\item \textbf{Confirmation rate.} The FALC period solutions and bootstrap tests confirm previously obtained periods for five objects, including asteroid (11219) Benbohn whose proposed period was already slightly above the spin barrier. Therefore, four objects are reliably identified as SFRs with periods between $1.06$ and $1.84$~hr. Seven candidates are conclusively rejected; three remain ambiguous pending additional data.

\item \textbf{Limitations of semi-dense survey photometry.}  
The four source surveys provide semi-dense cadences, yet most candidate periods drawn from their public catalogs prove incorrect or uncertain. These misidentifications demonstrate that sparse or semi-dense data alone cannot guarantee reliable period determinations, especially below the spin-barrier regime. Targeted, high‐cadence follow-up remains essential.

\item \textbf{Need for systematic validation.}  
Accurate rotation periods for fast rotators are inherently challenging because aliases concentrate near integer-hour periods, and small photometric amplitudes approach the noise floor. Routine validation campaigns—similar to ours—should accompany future survey releases to secure the integrity of period databases.

\item \textbf{Implications for the SFR population.}
Debiasing the LCDB candidate pool with our 29$\%$ confirmation efficiency yields a kilometre‑scale SFR fraction of $1.1\pm0.5\%$ ($1\sigma$). 
An alternative efficiency based only on secure periods gives a consistent $1.4\pm0.6\%$ result. These values overlap the $0.4$--$0.8\%$ range inferred from high‑cadence surveys, remaining compatible within the quoted uncertainties.

\end{enumerate}

Future wide-field facilities (e.g.\ the Rubin Observatory) will vastly increase the number of rotation-period measurements. However, the cadence bias highlighted here implies that complementary ground-based targeted campaigns will remain indispensable for vetting the shortest-period candidates and constraining the cohesive strength distribution of small asteroids.

\begin{acknowledgments}
BN acknowledges support by the Science Fund of the Republic of Serbia, GRANT No 7453, Demystifying enigmatic visitors of the near-Earth region (ENIGMA).
The authors also acknowledge financial support from project PID2021-126365NB-C21 (MCI/AEI/FEDER, UE) and from the Severo Ochoa grant CEX2021-001131-S funded by MCI/AEI/10.13039/501100011033.
Based on observations made at the Observatorio de Sierra Nevada (OSN), operated by the Instituto de Astrofísica de Andalucía (IAA-CSIC).
Part of the observations were obtained at the Astronomical Station Vidojevica, funded by the Ministry of Science, Technological Development and Innovation of the Republic of Serbia (Contract No. 451-03-136/2025-03/200002), and by the EC through the BELISSIMA project (FP7-REGPOT-2010-5, No. 256772).

\end{acknowledgments}

\bibliography{asteroids}{}

\begin{thebibliography}{}
\expandafter\ifx\csname natexlab\endcsname\relax\def\natexlab#1{#1}\fi
\providecommand{\url}[1]{\href{#1}{#1}}
\providecommand{\dodoi}[1]{doi:~\href{http://doi.org/#1}{\nolinkurl{#1}}}
\providecommand{\doeprint}[1]{\href{http://ascl.net/#1}{\nolinkurl{http://ascl.net/#1}}}
\providecommand{\doarXiv}[1]{\href{https://arxiv.org/abs/#1}{\nolinkurl{https://arxiv.org/abs/#1}}}

\bibitem[{B.~T. {Bolin} {et~al.}(2024){Bolin}, {Ghosal}, \&
  {Jedicke}}]{2024MNRAS.527.1633B}
{Bolin}, B.~T., {Ghosal}, M., \& {Jedicke}, R. 2024, \bibinfo{title}{{Rotation
  periods and colours of 10-m-scale near-Earth asteroids from CFHT target of
  opportunity streak photometry},} \mnras, 527, 1633,
  \dodoi{10.1093/mnras/stad3227}

\bibitem[{E. {Cameron}(2011){Cameron}}]{2011PASA...28..128C}
{Cameron}, E. 2011, \bibinfo{title}{{On the Estimation of Confidence Intervals
  for Binomial Population Proportions in Astronomy: The Simplicity and
  Superiority of the Bayesian Approach},} \pasa, 28, 128,
  \dodoi{10.1071/AS10046}

\bibitem[{A. {Cellino} {et~al.}(2024){Cellino}, {Tanga}, {Muinonen}, \&
  {Mignard}}]{2024A&A...687A.277C}
{Cellino}, A., {Tanga}, P., {Muinonen}, K., \& {Mignard}, F. 2024,
  \bibinfo{title}{{Asteroid spin and shape properties from Gaia DR3
  photometry},} \aap, 687, A277, \dodoi{10.1051/0004-6361/202449297}

\bibitem[{C.-K. {Chang} {et~al.}(2014){Chang}, {Ip}, {Lin}, {Cheng}, {Ngeow},
  {Yang}, {Waszczak}, {Kulkarni}, {Levitan}, {Sesar}, {Laher}, {Surace}, \&
  {Prince}}]{2014ApJ...788...17C}
{Chang}, C.-K., {Ip}, W.-H., {Lin}, H.-W., {et~al.} 2014, \bibinfo{title}{{313
  New Asteroid Rotation Periods from Palomar Transient Factory Observations},}
  \apj, 788, 17, \dodoi{10.1088/0004-637X/788/1/17}

\bibitem[{C.-K. {Chang} {et~al.}(2017){Chang}, {Lin}, {Ip}, {Lin}, {Kupfer},
  {Prince}, {Ye}, {Laher}, {Lee}, \& {Moon}}]{2017ApJ...840L..22C}
{Chang}, C.-K., {Lin}, H.-W., {Ip}, W.-H., {et~al.} 2017,
  \bibinfo{title}{{Confirmation of Large Super-fast Rotator (144977) 2005
  EC$_{127}$},} \apjl, 840, L22, \dodoi{10.3847/2041-8213/aa6ff5}

\bibitem[{C.-K. {Chang} {et~al.}(2019){Chang}, {Lin}, {Ip}, {Chen}, {Yeh},
  {Chambers}, {Magnier}, {Huber}, {Flewelling}, {Waters}, {Wainscoat}, \&
  {Schultz}}]{2019ApJS..241....6C}
{Chang}, C.-K., {Lin}, H.-W., {Ip}, W.-H., {et~al.} 2019,
  \bibinfo{title}{{Searching for Super-fast Rotators Using the Pan-STARRS 1},}
  \apjs, 241, 6, \dodoi{10.3847/1538-4365/ab01fe}

\bibitem[{C.-K. {Chang} {et~al.}(2022){Chang}, {Yeh}, {Tan}, {Ip}, {Kelley},
  {Ye}, {Lin}, {Ngeow}, {Bolin}, {Prince}, {Bellm}, {Dekany}, {Duev}, {Graham},
  \& {Zwicky Transient Facility Collaboration}}]{2022ApJ...932L...5C}
{Chang}, C.-K., {Yeh}, T.-S., {Tan}, H., {et~al.} 2022, \bibinfo{title}{{The
  Large Superfast Rotators Discovered by the Zwicky Transient Facility},}
  \apjl, 932, L5, \dodoi{10.3847/2041-8213/ac6e5e}

\bibitem[{V.~G. {Chiorny} {et~al.}(2025){Chiorny}, {Krugly}, {Shevchenko},
  {Slyusarev}, \& {Mikhalchenko}}]{2025P&SS..Chiorny}
{Chiorny}, V.~G., {Krugly}, Y.~N., {Shevchenko}, V.~G., {Slyusarev}, I.~G., \&
  {Mikhalchenko}, O. 2025, \bibinfo{title}{{Absolute photometry of small
  main-belt binary asteroids. Physical properties},} \planss, 262, 106118,
  \dodoi{10.1016/j.pss.2025.106118}

\bibitem[{W.-Y. {Dai} {et~al.}(2024){Dai}, {Yu}, {Cheng}, {Baoyin}, \&
  {Li}}]{2024A&A...684A.172D}
{Dai}, W.-Y., {Yu}, Y., {Cheng}, B., {Baoyin}, H., \& {Li}, J.-F. 2024,
  \bibinfo{title}{{Regolith resurfacing and shedding on spinning spheroidal
  asteroids: Dependence on the surface mechanical properties},} \aap, 684,
  A172, \dodoi{10.1051/0004-6361/202348112}

\bibitem[{M. {Delbo'} {et~al.}(2015){Delbo'}, {Mueller}, {Emery}, {Rozitis}, \&
  {Capria}}]{delbo-etal_2015}
{Delbo'}, M., {Mueller}, M., {Emery}, J.~P., {Rozitis}, B., \& {Capria}, M.~T.
  2015, {Asteroid Thermophysical Modeling} (University of Arizona Press),
  107--128

\bibitem[{M. {Delbo} {et~al.}(2014){Delbo}, {Libourel}, {Wilkerson}, {Murdoch},
  {Michel}, {Ramesh}, {Ganino}, {Verati}, \& {Marchi}}]{2014Natur.508..233D}
{Delbo}, M., {Libourel}, G., {Wilkerson}, J., {et~al.} 2014,
  \bibinfo{title}{{Thermal fatigue as the origin of regolith on small
  asteroids},} \nat, 508, 233, \dodoi{10.1038/nature13153}

\bibitem[{M. {Devog{\`e}le} {et~al.}(2024){Devog{\`e}le}, {Buzzi}, {Micheli},
  {Cano}, {Conversi}, {Jehin}, {Ferrais}, {Oca{\~n}a}, {F{\"o}hring}, {Drury},
  {Benkhaldoun}, \& {Jenniskens}}]{2024A&A...689A.211D}
{Devog{\`e}le}, M., {Buzzi}, L., {Micheli}, M., {et~al.} 2024,
  \bibinfo{title}{{Aperture photometry on asteroid trails: Detection of the
  fastest-rotating near-Earth object},} \aap, 689, A211,
  \dodoi{10.1051/0004-6361/202450263}

\bibitem[{N. {Erasmus} {et~al.}(2019){Erasmus}, {McNeill}, {Mommert},
  {Trilling}, {Sickafoose}, \& {Paterson}}]{2019ApJS..242...15E}
{Erasmus}, N., {McNeill}, A., {Mommert}, M., {et~al.} 2019, \bibinfo{title}{{A
  Taxonomic Study of Asteroid Families from KMTNET-SAAO Multiband Photometry},}
  \apjs, 242, 15, \dodoi{10.3847/1538-4365/ab1344}

\bibitem[{N. {Erasmus} {et~al.}(2018){Erasmus}, {McNeill}, {Mommert},
  {Trilling}, {Sickafoose}, \& {van Gend}}]{2018ApJS..237...19E}
{Erasmus}, N., {McNeill}, A., {Mommert}, M., {et~al.} 2018,
  \bibinfo{title}{{Taxonomy and Light-curve Data of 1000 Serendipitously
  Observed Main-belt Asteroids},} \apjs, 237, 19,
  \dodoi{10.3847/1538-4365/aac38f}

\bibitem[{P. {Fatka} {et~al.}(2025){Fatka}, {Pravec}, {Scheirich},
  {Ku{\v{s}}nir{\'a}k}, {Hornoch}, {Ku{\v{c}}{\'a}kov{\'a}}, {Ergashev}, {Souza
  de Joode}, {Burkhonov}, {Ehgamberdiev}, {Gal{\'a}d}, {Vil{\'a}gi}, {Reddy},
  {Dyvig}, {Ries}, {Snodgrass}, {Donaldson}, {Peixinho}, \&
  {Khalouei}}]{2025A&A...695A.139F}
{Fatka}, P., {Pravec}, P., {Scheirich}, P., {et~al.} 2025,
  \bibinfo{title}{{Spins and shapes of 11 near-Earth asteroids observed within
  the NEOROCKS project},} \aap, 695, A139, \dodoi{10.1051/0004-6361/202450027}

\bibitem[{M. {Fenucci} \& B. {Novakovi{\'c}}(2022){Fenucci} \&
  {Novakovi{\'c}}}]{2022SerAJ.204...51F}
{Fenucci}, M., \& {Novakovi{\'c}}, B. 2022, \bibinfo{title}{{MERCURY and ORBFIT
  Packages for Numerical Integration of Planetary Systems: Implementation of
  the Yarkovsky and YORP Effects},} Serbian Astronomical Journal, 204, 51,
  \dodoi{10.2298/SAJ2204051F}

\bibitem[{M. {Fenucci} {et~al.}(2023){Fenucci}, {Novakovi{\'c}}, \&
  {Mar{\v{c}}eta}}]{2023A&A...675A.134F}
{Fenucci}, M., {Novakovi{\'c}}, B., \& {Mar{\v{c}}eta}, D. 2023,
  \bibinfo{title}{{The low surface thermal inertia of the rapidly rotating
  near-Earth asteroid 2016 GE1},} \aap, 675, A134,
  \dodoi{10.1051/0004-6361/202346160}

\bibitem[{M. {Fenucci} {et~al.}(2021){Fenucci}, {Novakovi{\'c}},
  {Vokrouhlick{\'y}}, \& {Weryk}}]{2021A&A...647A..61F}
{Fenucci}, M., {Novakovi{\'c}}, B., {Vokrouhlick{\'y}}, D., \& {Weryk}, R.~J.
  2021, \bibinfo{title}{{Low thermal conductivity of the superfast rotator
  (499998) 2011 PT},} \aap, 647, A61, \dodoi{10.1051/0004-6361/202039628}

\bibitem[{I. {Fodde} \& F. {Ferrari}(2025){Fodde} \&
  {Ferrari}}]{2025A&A...695A..30F}
{Fodde}, I., \& {Ferrari}, F. 2025, \bibinfo{title}{{Dynamical modelling of
  rubble pile asteroids using data-driven techniques},} \aap, 695, A30,
  \dodoi{10.1051/0004-6361/202452432}

\bibitem[{M. {Gowanlock} {et~al.}(2024){Gowanlock}, {Trilling}, {McNeill},
  {Kramer}, \& {Chernyavskaya}}]{2024AJ....168..181G}
{Gowanlock}, M., {Trilling}, D.~E., {McNeill}, A., {Kramer}, D., \&
  {Chernyavskaya}, M. 2024, \bibinfo{title}{{Asteroid Period Solutions from
  Combined Dense and Sparse Photometry},} \aj, 168, 181,
  \dodoi{10.3847/1538-3881/ad6cdd}

\bibitem[{J. {Hanu{\v{s}}} {et~al.}(2018){Hanu{\v{s}}}, {Delbo'},
  {{\v{D}}urech}, \& {Al{\'\i}-Lagoa}}]{2018Icar..309..297H}
{Hanu{\v{s}}}, J., {Delbo'}, M., {{\v{D}}urech}, J., \& {Al{\'\i}-Lagoa}, V.
  2018, \bibinfo{title}{{Thermophysical modeling of main-belt asteroids from
  WISE thermal data},} \icarus, 309, 297, \dodoi{10.1016/j.icarus.2018.03.016}

\bibitem[{A.~W. {Harris} {et~al.}(1989){Harris}, {Young}, {Bowell}, {Martin},
  {Millis}, {Poutanen}, {Scaltriti}, {Zappala}, {Schober}, {Debehogne}, \&
  {Zeigler}}]{1989Icar...77..171H}
{Harris}, A.~W., {Young}, J.~W., {Bowell}, E., {et~al.} 1989,
  \bibinfo{title}{{Photoelectric observations of asteroids 3, 24, 60, 261, and
  863},} \icarus, 77, 171, \dodoi{10.1016/0019-1035(89)90015-8}

\bibitem[{A.~N. {Heinze} {et~al.}(2018){Heinze}, {Tonry}, {Denneau},
  {Flewelling}, {Stalder}, {Rest}, {Smith}, {Smartt}, \&
  {Weiland}}]{2018AJ....156..241H}
{Heinze}, A.~N., {Tonry}, J.~L., {Denneau}, L., {et~al.} 2018,
  \bibinfo{title}{{A First Catalog of Variable Stars Measured by the Asteroid
  Terrestrial-impact Last Alert System (ATLAS)},} \aj, 156, 241,
  \dodoi{10.3847/1538-3881/aae47f}

\bibitem[{M. {Hirabayashi} \& D.~J. {Scheeres}(2019){Hirabayashi} \&
  {Scheeres}}]{2019Icar..317..354H}
{Hirabayashi}, M., \& {Scheeres}, D.~J. 2019, \bibinfo{title}{{Rotationally
  induced failure of irregularly shaped asteroids},} \icarus, 317, 354,
  \dodoi{10.1016/j.icarus.2018.08.003}

\bibitem[{S. {Hu} {et~al.}(2021){Hu}, {Richardson}, {Zhang}, \&
  {Ji}}]{2021MNRAS.502.5277H}
{Hu}, S., {Richardson}, D.~C., {Zhang}, Y., \& {Ji}, J. 2021,
  \bibinfo{title}{{Critical spin periods of sub-km-sized cohesive rubble-pile
  asteroids: dependences on material parameters},} \mnras, 502, 5277,
  \dodoi{10.1093/mnras/stab412}

\bibitem[{D. {Hung} {et~al.}(2022){Hung}, {Hanu{\v{s}}}, {Masiero}, \&
  {Tholen}}]{2022PSJ.....3...56H}
{Hung}, D., {Hanu{\v{s}}}, J., {Masiero}, J.~R., \& {Tholen}, D.~J. 2022,
  \bibinfo{title}{{Thermal Properties of 1847 WISE-observed Asteroids},} \psj,
  3, 56, \dodoi{10.3847/PSJ/ac4d1f}

\bibitem[{D. {Kramer} {et~al.}(2023){Kramer}, {Gowanlock}, {Trilling},
  {McNeill}, \& {Erasmus}}]{2023A&C....4400711K}
{Kramer}, D., {Gowanlock}, M., {Trilling}, D., {McNeill}, A., \& {Erasmus}, N.
  2023, \bibinfo{title}{{Removing aliases in time-series photometry},}
  Astronomy and Computing, 44, 100711, \dodoi{10.1016/j.ascom.2023.100711}

\bibitem[{J. {Licandro} {et~al.}(2023){Licandro}, {Popescu}, {Tatsumi},
  {Alarcon}, {Serra-Ricart}, {Medeiros}, {Morate}, {Tinaut-Ruano}, \& {de
  Le{\'o}n}}]{2023MNRAS.521.3784L}
{Licandro}, J., {Popescu}, M., {Tatsumi}, E., {et~al.} 2023,
  \bibinfo{title}{{Observations of two superfast rotator NEAs: 2021 NY$_{1}$
  and 2022 AB},} \mnras, 521, 3784, \dodoi{10.1093/mnras/stad708}

\bibitem[{F. {Monteiro} {et~al.}(2020){Monteiro}, {Silva}, {Tamayo},
  {Rodrigues}, \& {Lazzaro}}]{2020MNRAS.495.3990M}
{Monteiro}, F., {Silva}, J.~S., {Tamayo}, F., {Rodrigues}, T., \& {Lazzaro}, D.
  2020, \bibinfo{title}{{Shape model and spin direction analysis of PHA
  (436724) 2011 UW158: a large superfast rotator},} \mnras, 495, 3990,
  \dodoi{10.1093/mnras/staa1401}

\bibitem[{B. {Novakovi{\'c}} {et~al.}(2024){Novakovi{\'c}}, {Fenucci},
  {Mar{\v{c}}eta}, \& {Pavela}}]{2024PSJ.....5...11N}
{Novakovi{\'c}}, B., {Fenucci}, M., {Mar{\v{c}}eta}, D., \& {Pavela}, D. 2024,
  \bibinfo{title}{{ASTERIA-Asteroid Thermal Inertia Analyzer},} \psj, 5, 11,
  \dodoi{10.3847/PSJ/ad08c0}

\bibitem[{B. {Novakovi{\'c}} {et~al.}(2022){Novakovi{\'c}}, {Vokrouhlick{\'y}},
  {Spoto}, \& {Nesvorn{\'y}}}]{2022CeMDA.134...34N}
{Novakovi{\'c}}, B., {Vokrouhlick{\'y}}, D., {Spoto}, F., \& {Nesvorn{\'y}}, D.
  2022, \bibinfo{title}{{Asteroid families: properties, recent advances, and
  future opportunities},} Celestial Mechanics and Dynamical Astronomy, 134, 34,
  \dodoi{10.1007/s10569-022-10091-7}

\bibitem[{A. {P{\'a}l} {et~al.}(2020){P{\'a}l}, {Szak{\'a}ts}, {Kiss},
  {B{\'o}di}, {Bogn{\'a}r}, {Kalup}, {Kiss}, {Marton}, {Moln{\'a}r}, {Plachy},
  {S{\'a}rneczky}, {Szab{\'o}}, \& {Szab{\'o}}}]{2020ApJS..247...26P}
{P{\'a}l}, A., {Szak{\'a}ts}, R., {Kiss}, C., {et~al.} 2020,
  \bibinfo{title}{{Solar System Objects Observed with TESS{\textemdash}First
  Data Release: Bright Main-belt and Trojan Asteroids from the Southern
  Survey},} \apjs, 247, 26, \dodoi{10.3847/1538-4365/ab64f0}

\bibitem[{D. {Parrott}(2020){Parrott}}]{2020JAVSO..48..262P}
{Parrott}, D. 2020, \bibinfo{title}{{Tycho Tracker: A New Tool to Facilitate
  the Discovery and Recovery of Asteroids Using Synthetic Tracking and Modern
  GPU Hardware (Abstract)},} \jaavso, 48, 262

\bibitem[{D. {Polishook}(2013){Polishook}}]{2013MPBu...40...42P}
{Polishook}, D. 2013, \bibinfo{title}{{Fast Rotation of the NEA 2012 TC4
  Indicates a Monolithic Structure},} Minor Planet Bulletin, 40, 42

\bibitem[{P. {Pravec} \& A.~W. {Harris}(2000){Pravec} \&
  {Harris}}]{2000Icar..148...12P}
{Pravec}, P., \& {Harris}, A.~W. 2000, \bibinfo{title}{{Fast and Slow Rotation
  of Asteroids},} \icarus, 148, 12, \dodoi{10.1006/icar.2000.6482}

\bibitem[{E. {Rond{\'o}n} {et~al.}(2020){Rond{\'o}n}, {Lazzaro}, {Rodrigues},
  {Carvano}, {Roig}, {Monteiro}, {Arcoverde}, {Medeiros}, {Silva}, {Jasmim},
  {Pr{\'a}}, {Hasselmann}, {Ribeiro}, {D{\'a}valos}, \&
  {Souza}}]{2020PASP..132f5001R}
{Rond{\'o}n}, E., {Lazzaro}, D., {Rodrigues}, T., {et~al.} 2020,
  \bibinfo{title}{{OASI: A Brazilian Observatory Dedicated to the Study of
  Small Solar System Bodies{\textemdash}Some Results on NEO's Physical
  Properties},} \pasp, 132, 065001, \dodoi{10.1088/1538-3873/ab87a7}

\bibitem[{B. {Rozitis} {et~al.}(2014){Rozitis}, {Maclennan}, \&
  {Emery}}]{2014Natur.512..174R}
{Rozitis}, B., {Maclennan}, E., \& {Emery}, J.~P. 2014,
  \bibinfo{title}{{Cohesive forces prevent the rotational breakup of
  rubble-pile asteroid (29075) 1950 DA},} \nat, 512, 174,
  \dodoi{10.1038/nature13632}

\bibitem[{P. {S{\'a}nchez} \& D.~J. {Scheeres}(2020){S{\'a}nchez} \&
  {Scheeres}}]{2020Icar..33813443S}
{S{\'a}nchez}, P., \& {Scheeres}, D.~J. 2020, \bibinfo{title}{{Cohesive
  regolith on fast rotating asteroids},} \icarus, 338, 113443,
  \dodoi{10.1016/j.icarus.2019.113443}

\bibitem[{R. {Strauss} {et~al.}(2024){Strauss}, {McNeill}, {Trilling},
  {Valdes}, {Bernardinelli}, {Fuentes}, {Gerdes}, {Holman}, {Juri{\'c}}, {Lin},
  {Markwardt}, {Mommert}, {Napier}, {Oldroyd}, {Payne}, {Rivkin},
  {Schlichting}, {Sheppard}, {Smotherman}, {Trujillo}, {Adams}, \&
  {Chandler}}]{2024AJ....168..184S}
{Strauss}, R., {McNeill}, A., {Trilling}, D.~E., {et~al.} 2024,
  \bibinfo{title}{{The DECam Ecliptic Exploration Project (DEEP). VII. The
  Strengths of Three Superfast Rotating Main-belt Asteroids from a Preliminary
  Search of DEEP Data},} \aj, 168, 184, \dodoi{10.3847/1538-3881/ad7366}

\bibitem[{N. {Tak{\'a}cs} {et~al.}(2025){Tak{\'a}cs}, {Kiss}, {Szak{\'a}ts},
  {Plachy}, {Kalup}, {Szab{\'o}}, {Moln{\'a}r}, {S{\'a}rneczky}, {Szab{\'o}},
  {B{\'o}di}, \& {P{\'a}l}}]{2025ApJ...986L..33T}
{Tak{\'a}cs}, N., {Kiss}, C., {Szak{\'a}ts}, R., {et~al.} 2025,
  \bibinfo{title}{{Three Fast-spinning Medium-sized Hilda Asteroids Uncovered
  by TESS},} \apjl, 986, L33, \dodoi{10.3847/2041-8213/ade05b}

\bibitem[{P. {Tanga} {et~al.}(2023){Tanga}, {Pauwels}, {Mignard}, {Muinonen},
  {Cellino}, {David}, {Hestroffer}, {Spoto}, {Berthier}, {Guiraud}, {Roux},
  {Carry}, {Delbo}, {Dell'Oro}, {Fouron}, {Galluccio}, {Jonckheere}, {Klioner},
  {Lefustec}, {Liberato}, {Ord{\'e}novic}, {Oreshina-Slezak}, {Penttil{\"a}},
  {Pailler}, {Panem}, {Petit}, {Portell}, {Poujoulet}, {Thuillot}, {Van
  Hemelryck}, {Burlacu}, {Lasne}, \& {Managau}}]{2023A&A...674A..12T}
{Tanga}, P., {Pauwels}, T., {Mignard}, F., {et~al.} 2023, \bibinfo{title}{{Gaia
  Data Release 3. The Solar System survey},} \aap, 674, A12,
  \dodoi{10.1051/0004-6361/202243796}

\bibitem[{J.~L. {Tonry} {et~al.}(2018){Tonry}, {Denneau}, {Flewelling},
  {Heinze}, {Onken}, {Smartt}, {Stalder}, {Weiland}, \&
  {Wolf}}]{2018ApJ...867..105T}
{Tonry}, J.~L., {Denneau}, L., {Flewelling}, H., {et~al.} 2018,
  \bibinfo{title}{{The ATLAS All-Sky Stellar Reference Catalog},} \apj, 867,
  105, \dodoi{10.3847/1538-4357/aae386}

\bibitem[{O. {Vaduvescu} {et~al.}(2022){Vaduvescu}, {Aznar Macias}, {Wilson},
  {Zegmott}, {P{\'e}rez Toledo}, {Predatu}, {Gherase}, {Pinter}, {Pozo Nunez},
  {Ulaczyk}, {Soszy{\'n}ski}, {Mr{\'o}z}, {Wrona}, {Iwanek}, {Szymanski},
  {Udalski}, {Char}, {Salas Olave}, {Aravena-Rojas}, {Vergara}, {Saez},
  {Unda-Sanzana}, {Alcalde}, {de Burgos}, {Nespral}, {Galera-Rosillo}, {Amos},
  {Hibbert}, {L{\'o}pez-Comazzi}, {Oey}, {Serra-Ricart}, {Licandro}, \&
  {Popescu}}]{2022EM&P..126....6V}
{Vaduvescu}, O., {Aznar Macias}, A., {Wilson}, T.~G., {et~al.} 2022,
  \bibinfo{title}{{The EURONEAR Lightcurve Survey of Near Earth Asteroids
  2017{\textendash}2020},} Earth Moon and Planets, 126, 6,
  \dodoi{10.1007/s11038-022-09548-4}

\bibitem[{D.~E. {Vavilov} \& B. {Carry}(2025){Vavilov} \&
  {Carry}}]{2025A&A...693A..66V}
{Vavilov}, D.~E., \& {Carry}, B. 2025, \bibinfo{title}{{Rotation periods of
  asteroids from light curves of TESS data},} \aap, 693, A66,
  \dodoi{10.1051/0004-6361/202348940}

\bibitem[{D. {Vokrouhlick{\'y}} {et~al.}(2015){Vokrouhlick{\'y}}, {Bottke},
  {Chesley}, {Scheeres}, \& {Statler}}]{2015aste.book..509V}
{Vokrouhlick{\'y}}, D., {Bottke}, W.~F., {Chesley}, S.~R., {Scheeres}, D.~J.,
  \& {Statler}, T.~S. 2015, {The Yarkovsky and YORP Effects} (University of
  Arizona Press), 509--531, \dodoi{10.2458/azu\_uapress\_9780816532131-ch027}

\bibitem[{K.~J. {Walsh}(2018){Walsh}}]{2018ARA&A..56..593W}
{Walsh}, K.~J. 2018, \bibinfo{title}{{Rubble Pile Asteroids},} \araa, 56, 593,
  \dodoi{10.1146/annurev-astro-081817-052013}

\bibitem[{B.~D. {Warner} \& A.~W. {Harris}(2011){Warner} \&
  {Harris}}]{2011Icar..216..610W}
{Warner}, B.~D., \& {Harris}, A.~W. 2011, \bibinfo{title}{{Using sparse
  photometric data sets for asteroid lightcurve studies},} \icarus, 216, 610,
  \dodoi{10.1016/j.icarus.2011.10.007}

\bibitem[{T.-S. {Yeh} {et~al.}(2020){Yeh}, {Li}, {Chang}, {Zhao}, {Ji}, {Lin},
  \& {Ip}}]{2020AJ....160...73Y}
{Yeh}, T.-S., {Li}, B., {Chang}, C.-K., {et~al.} 2020, \bibinfo{title}{{The
  Asteroid Rotation Period Survey Using the China Near-Earth Object Survey
  Telescope (CNEOST)},} \aj, 160, 73, \dodoi{10.3847/1538-3881/ab9a32}

\end{thebibliography}
\bibliographystyle{aasjournalv7}



\end{document}